\documentclass[fleqn,usenatbib]{mnras}

\usepackage{newtxtext,newtxmath}
\usepackage{xcolor}

\usepackage[T1]{fontenc}
\usepackage{ae,aecompl}
\usepackage{hyperref}

\usepackage{graphicx}	
\usepackage{amsmath}	
\usepackage{amssymb}	

\title[The X-Shooter spectrum of Makemake]{The Dwarf Planet Makemake as seen by X-Shooter}

\author[A. Alvarez-Candal et al.]
{Alvaro Alvarez-Candal,$^{1,2,3}$\thanks{E-mail: alvaro.alvarez@ua.es, Incoming Severo Ochoa visitor at the IAA-CSIC.}
Ana Carolina Souza-Feliciano,$^{2,4}$
Walter Martins-Filho,$^{2}$
\newauthor
Noem\'i Pinilla-Alonso,$^{4,5}$
and 
Jos\'e Luis Ortiz$^{3}$
\\
$^{1}$IUFACyT, Universidad de Alicante, San Vicent del Raspeig, E03080, Alicante, Spain\\
$^{2}$Observat\'orio Nacional / MCTIC, Rua General Jos\'e Cristino 77, Rio de Janeiro, RJ, 20921-400, Brazil\\
$^3$Instituto de Astrof\'isica de Andaluc\'ia, CSIC, Apt 3004, E18080 Granada, Spain\\
$^4$Florida Space Institute, University of Central Florida, FL, USA\\
$^5$Arecibo Observatory, University of Central Florida, HC-3 Box 53995, Arecibo 00612, Puerto Rico
}

\date{Accepted XXX. Received YYY; in original form ZZZ}

\pubyear{2020}

\begin{document}
\label{firstpage}
\pagerange{\pageref{firstpage}--\pageref{lastpage}}
\maketitle

\begin{abstract}
Makemake is one of the brightest known trans-Neptunian objects, as such, it has been widely observed. Nevertheless, its visible to near-infrared spectrum has not been completely observed in medium resolving power, aimed at studying in detail the absorption features of CH$_4$ ice. In this paper we present the spectrum of Makemake observed with X-Shooter at the Very Large Telescope (Chile). We analyse the detected features, measuring their location and depth. Furthermore, we compare Makemake's spectrum with that of Eris, obtained with the same instrument and similar setup, to conclude that the bands of the CH$_4$ ice in both objects show similar shifts.
\end{abstract}

\begin{keywords}
methods: observational -- 
techniques: spectroscopic -- 
Kuiper belt: Makemake
\end{keywords}


\section{Introduction}\label{sec:introduction}
(136472) Makemake (just Makemake in the remaining of the text) is one of the largest trans-Neptunian objects (TNOs), having an equivalent diameter of {($1,430\pm9$)}\footnote{All errors shown in this paper correspond to 1$\sigma$ confidence level.} km and a visible albedo of {$(0.77\pm0.03)$} \citep{ortiz12}. Its rotational period is estimated as {$(22.8266\pm0.001)$} h, double peaked light-curve, with a peak-to-peak amplitude of {$(0.032\pm0.005)$} mags \citep{hromakina2019}, in contrast to the value of {$(7.7710\pm0.0030)$} h (single peaked light-curve) reported by \cite{heinz09}.

Recently, \cite{parke16} reported the detection of a satellite about 8 magnitudes fainter than Makemake, whose orbit remains yet to be determined. This orbit will allow to measure the mass of Makemake and, therefore, its density, which can be contrasted to different values in the literature: $(1.7\pm0.3)$ gcm$^{-3}$ from \cite{ortiz12}, > 1.98 gcm$^{-3}$ from \cite{brown13}, or $\lesssim1.8$ gcm$^{-3}$ estimated by \cite{bierson2019}. Noteworthy, \cite{hromakina2019} also suggested the existence of a second satellite, close to the primary, whose existence remains to be confirmed.

The visible to near-infrared (VNIR) spectrum of Makemake is dominated by absorption bands of CH$_{4}$ ice \citep{lican06}. Their central positions are very close to the ones of pure CH$_{4}$ ice measured on laboratory, but still slightly blue-shifted, which seem to indicate that Makemake is one the objects with largest amount of CH$_4$ ice on its  surface. {These shifts are attributed to the mixture of ices: CH$_4$ and possibly N$_2$ in different levels of dilution \citep[for instance, see][]{quiri97,brunetto2008,protopapa2015}}. For the sake of comparison, Pluto \citep[and references therein]{merli10} presents large blue-shifts (usually > 10 \AA) in the spectral signatures of CH$_4$, interpreted as larger dilutions of CH$_4$ in N$_2$. {Although other ices might be present on the surface of Makemake, for instance CO, as observed in Pluto, its effect might be negligible \citep{tan2018}. }

Several authors have presented VNIR spectroscopy of Makemake, measuring the position and depth of the CH$_4$ absorption features {\citep[for example][]{lican06,tegle007,tegler2008,loren15,perna17}}. The deep and broad absorption features (see Fig. \ref{fig01}) are interpreted in terms of very large slabs formed by sintering \citep{elusz07}. The small blue-shifts of the absorption features are in agreement with the \cite{schal07} \citep[updated in][]{brown2012AREPS} model of volatile retention that shows Makemake on the border of N$_2$ ice retention region. Nevertheless, this model should be interpreted with care because it only gives an upper limit on the survivability of the ices, {furthermore, the Jeans escape rate has been shown to be even smaller than expected for Pluto's atmosphere \citep[see][and references therein]{Young2020}}. 

Most of the measurements of the relative position of the CH$_4$ ice bands of Makemake were performed in the visible range $(\lambda<0.9$ $\mu$m) {with resolving power usually $\leq1,000$, with a few exceptions, e.g., $\sim1,400$ in \cite{tegler2008} and about 3,000 in \cite{perna17}.
Before continuing it is important to mention that the resolving power is the ability of a spectrograph to resolve a spectral line of width at half maximum $\delta\lambda$ at a wavelength $\lambda$ and is given by $R=\lambda/\delta\lambda$. In the NIR, most of the published spectra have $R<100$, with the exception of \cite{brown15}'s with {$R\sim2,500$}.} Due to the usually low resolving power of the NIR spectra, it is not possible to carry on detailed studies of absorption features above 1.0 $\mu$m. 
To stress the importance of mid-resolving power NIR spectroscopy {($1,000\lesssim R\lesssim10,000$)}, \cite{brown15} showed that irradiation products of CH$_4$ ice, for instance C$_2$H$_6$ ice, improve the spectral modelling of Makemake, especially for $\lambda>1.6$ $\mu$m. Therefore, in this work we present mid-resolving power spectrum of Makemake {($R>4,000$)} in a large spectral range. The spectrum was obtained with X-Shooter @ Very Large Telescope {(VLT)} and it has the highest SNR obtained so far simultaneously in the VNIR range.

This work is organised as follows, in the next section we describe the observations and data reduction. In Sect. 3 we show our results, while in Sect. 4 we present the discussion and in Sect. 5 the conclusions drawn from this work.

\section{Observations and data reduction}

Makemake was observed in service mode using X-Shooter\footnote{\tt \url{https://www.eso.org/sci/facilities/paranal/instruments/xshooter.html}}, attached to the Cassegrain focus of the unit 2 of the VLT (Cerro Paranal, Chile) on April 26, 2013. X-Shooter is an echelle spectrograph able to obtain at once a complete spectrum between 0.35 and 2.4 $\mu$m. The incoming beam of light is split using two {dichroic beam splitters} and redirected into three arms: UVB ($0.35-0.55$ $\mu$m), VIS ($0.55-1.0$ $\mu$m), and NIR ($1.0-2.4$ $\mu$m). Each arm works as an independent spectrograph recording a different part of the spectrum.

We used the instrumental setup shown in Table \ref{table:1}. In the NIR arm we selected a restricted mode that only records the spectrum up to 1.8 $\mu$m, while everything above is blocked out. The blocked region shows a very strong thermal contamination that damages the signal for faint objects, even contaminating the H region. Using this setup we obtain an increased efficiency in the $1.5-1.8$ $\mu$m region. The observations were taken nodding on the slit following a standard ABBA pattern.
\begin{table*}
	\centering
	\caption{Observational Circumstances}\label{table:1}
	\begin{tabular}{c c c c c c }
\hline
Object& Arm & Read Out Mode / binning & Slit (arcsec) & Exptime (s) & Airmass \\
\hline
         & UVB & 100 khz / $1\times2$ & 1.0 & $4\times480$ &       \\
Makemake & VIS & 100 khz / $1\times2$ & 0.9 & $4\times500$ & 1.761 \\
         & NIR & Non Destructive Mode & 0.9 & $4\times480$ &       \\
\hline
         & UVB & 100 khz / $1\times2$ & 1.0 & $2\times0.7$ &       \\
HD89010  & VIS & 100 khz / $1\times2$ & 0.9 & $2\times0.7$ & 1.550 \\
         & NIR & Non Destructive Mode & 0.9 & $2\times0.7$ &       \\
\hline
\end{tabular}
\end{table*}
We used the star HD89010, a.k.a. 35 Leo, of spectral type G1.5IV-V, to remove the solar signature from Makemake's spectrum and as telluric star (see Table \ref{table:1}).

The data were delivered via ftp package including all necessary calibration files and reduced using the X-Shooter pipeline version 2.0.0, {which handles the whole reduction process}: BIAS or DARK correction, whichever is necessary, FLAT-FIELDING, order identification, rectification of the raw spectra, wavelength calibration, sky-subtraction, and merging of the orders into, first, a two-dimensional image, and then into a one-dimensional spectrum. We did not use this last spectrum, opting to make our own extraction form the 2D image using {\tt apall.iraf}. This ensures a better control over the final signal-to-noise ratio of the spectrum.

After extraction of the spectra we divided the spectrum of Makemake by that of the star removing the solar signature and correcting most of the telluric absorption due to Earth's atmosphere, obtaining our final reflectance spectrum. Finally, we normalised it to unity at 0.55 $\mu$m.

\subsection{Filtering}

The resulting reflectance spectrum is still noisy, which makes {it} difficult to detect where an absorption feature starts and where it ends. It also has many remaining bad pixels that were not flagged during the pipeline processing. We decided to filter the spectrum using {\it wavelets}, in particular the family of wavelets {\it Coiflet} as they showed an optimal behaviour in comparison with other families of wavelets \cite[see][]{souza18}. We also chose to work with wavelets instead of other filtering techniques, such as re-binning, running box, or Fourier analysis because it revealed as the technique that better respected the input signal \citep{souza18}.

Wavelets de-construct the signal into two parts: one principal and one residual. The deconvolution occurs simultaneously in the spatial and frequency domains \citep{starc06}. We used scale of 2 because we only wanted to remove bad pixels. We used the hard-filtering that removes coefficients below a certain threshold, respecting the shape of the features. 

\begin{figure}
	\includegraphics[width=\columnwidth]{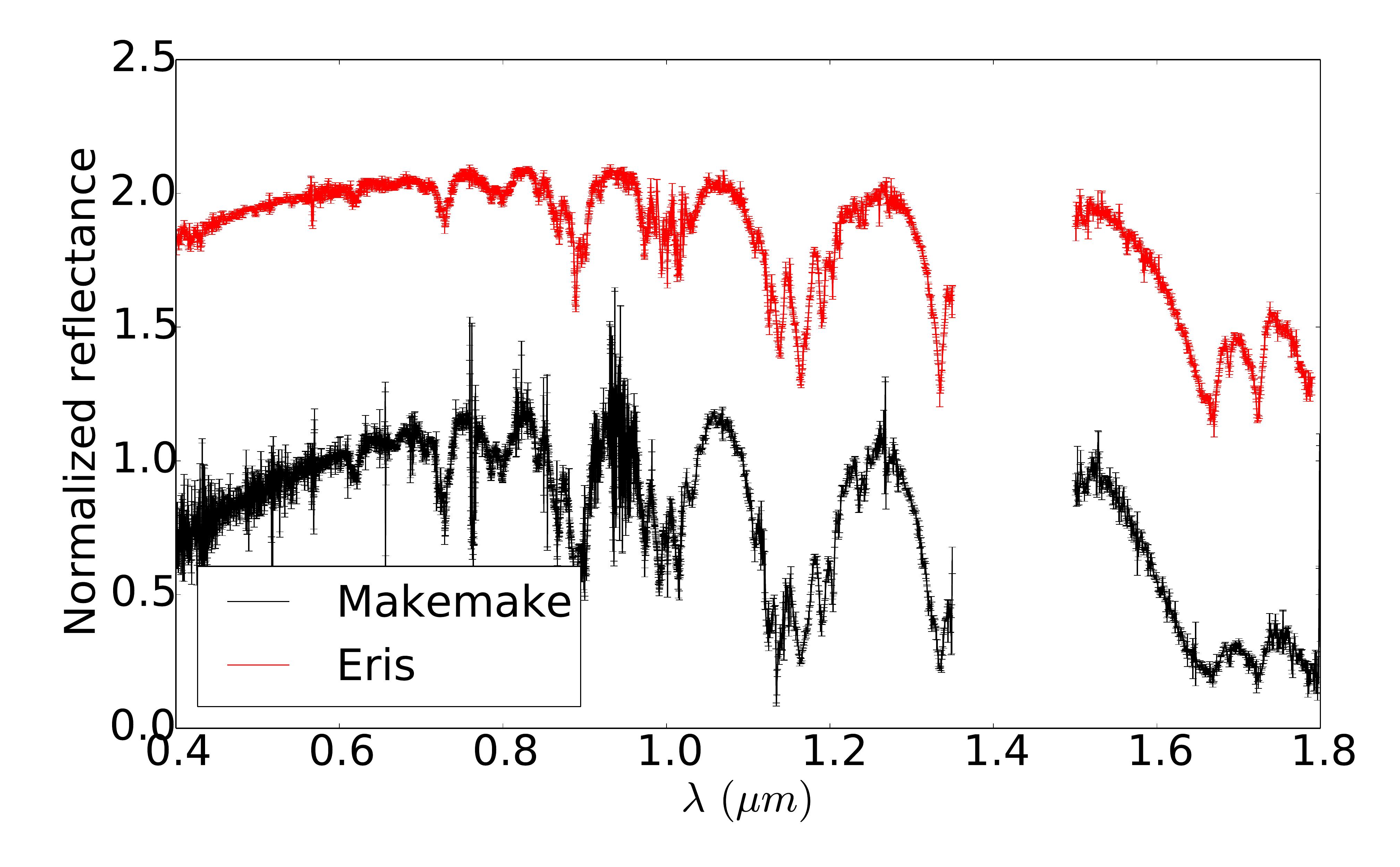}
    \caption{Spectra of Makemake, in black, and Eris, in red. Both spectra are normalised to 0.55 $\mu$m, while Eris' was shifted in the y-scale for clarity. The region between 1.4 and 1.5 $\mu$m was removed due to strong atmospheric absorption. Both spectra were obtained with X-Shooter.}
    \label{fig01}
\end{figure}

Figure \ref{fig01} shows the spectrum of Makemake after the process (bottom spectrum). It is possible to see a few remaining bad pixels, which occur at the joint of  the different echelle orders, especially in the VIS arm. We decided to not go further with the cleaning process to avoid the destruction of real features of the spectrum. {We also discarded the region below 0.4 $\mu$m because it was too noisy to be reliable.}
We will use this spectrum in the remaining of the work.

\section{Results}

In this work we will use as comparison a spectrum of Eris obtained also with X-Shooter because we aim to compare spectra of similar resolving power, obtained with the same instrument. The details of Eris' observation and data reduction can be found in \cite{alcan11}. Note that the spectrum of Eris was de-noised in the same way as described above.
To make the most of the resolving power of our spectrum we will analyse first the characteristics of the absorption features, i.e., band depth ($D$) and wavelength shifts ($\Delta\lambda$), and then look into the possible surface composition.

\subsection{Spectrum in the visible}
An important characteristic of the spectrum is its colour in the visible range. {Therefore, we first compared the visible spectral slope, $S'$, using the CANA\footnote{The CANA toolkit (Codes for ANalysis of Asteroids) is a Python package specifically developed to facilitate the study of features in asteroids spectroscopic and spectrophotometric data.} package \citep{cana}, that fits a linear function between 0.4 and 0.52 $\mu$m and estimates this parameter. The spectral slope measured for Makemake is $S_M' = (21.2 \pm 0.6$) $\%/1000$ \AA\ and for Eris is $S_E' = (13.5 \pm 0.2$) $\%/1000$ \AA.} Notice that these values are not to be {directly compared with others in the literature \cite[for instance][]{lorenzi2016} because these values are dependent of the exact definition used for $S'$, which changes from work to work.} 

{A direct comparison can be made using $(B-V)$ and $(V-R)$ colours. We used the transmission curves, $T(\lambda)$, of the three filters\footnote{http://svo2.cab.inta-csic.es/theory/fps/index.php?mode=browse\&gname=Generic} to weight the spectra and obtain relative magnitudes, which were then transformed into standard magnitudes using solar colours \cite[from][]{ramirez2012} following
$$(M_1 - M_2)_{\rm obj}= -2.5\log{(f_1/f_2)}+({M_1-M_2})_\odot,$$
where $$f_i=\frac{\int{T_i(\lambda})flux(\lambda)d\lambda}{\int{T_i(\lambda})d\lambda},~i=B,V,R.$$
We obtained, for Makemake, $(B-V) = 0.868\pm0.004$ and $(V-R) = 0.449\pm0.003$, while for Eris $(B-V) = 0.782\pm0.003$ and $(V-R) = 0.393\pm0.003$. These values are in agreement with those reported in the MBOSS database by \cite{mboss}: $(B-V) = 0.84\pm0.02$ and $(V-R) = 0.48\pm0.02$ for Makemake, and $(B-V) = 0.78\pm0.03$ and $(V-R) = 0.39\pm0.05$ for Eris.}

The most accepted hypothesis to explain the red colours of Makemake and Eris is the existence of complex organics molecules (tholins) formed from simple organics by photolysis \citep[e.g.][]{simonia2018, khare84}. {A possible explanation for the difference could be that Eris has less tholins than Makemake}.

\subsubsection{Subtle features}

\begin{figure}
 \centering
 \includegraphics[width=\columnwidth]{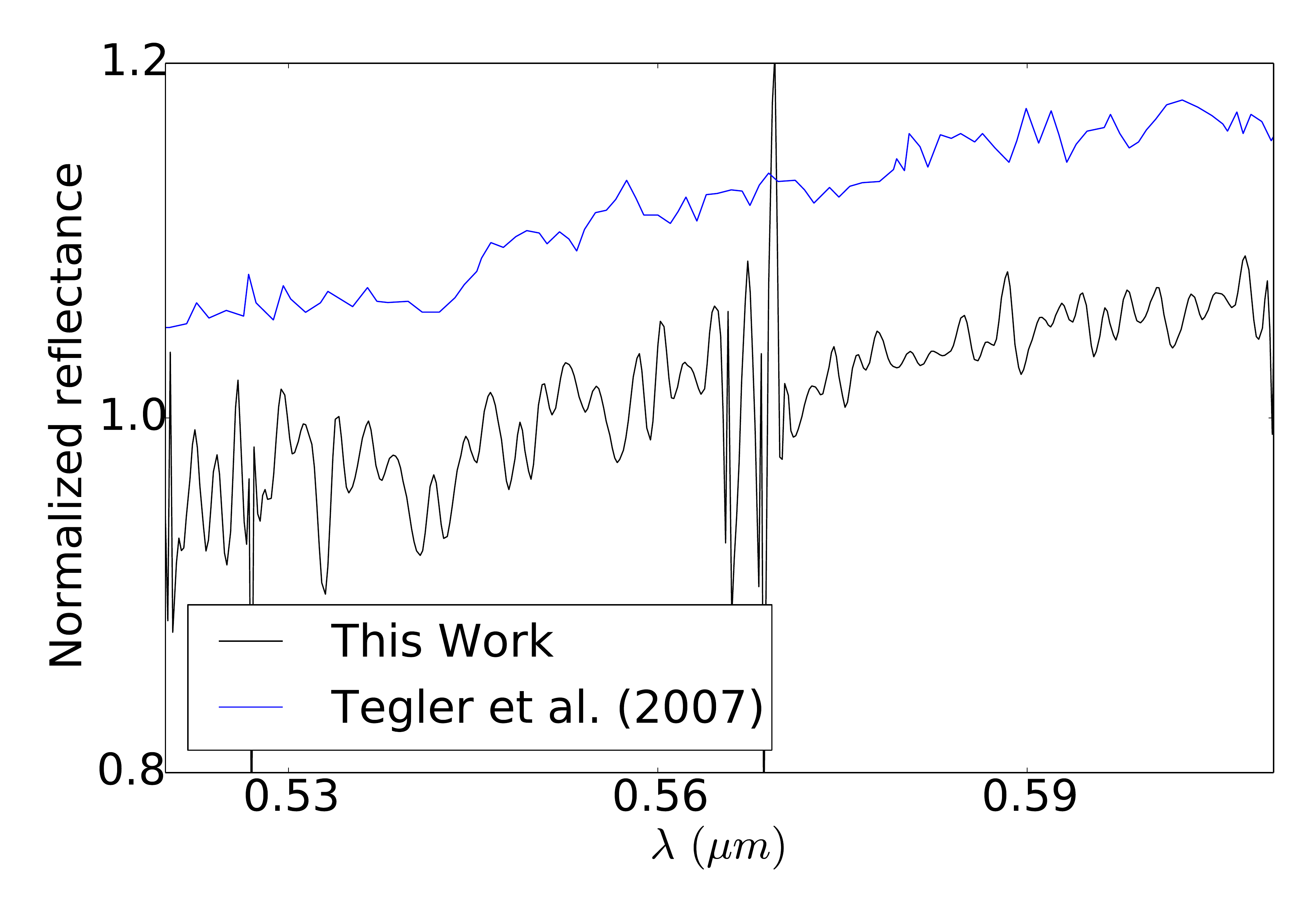}
 \caption{Comparison between spectra of Makemake zoomed to the 0.52 $\mu$m and 0.61 $\mu$m. Both were normalised to unite at 0.55 $\mu$m, Tegler et al.'s is shifted by 0.1 in the y-axis for clarity.}\label{figcomp}
\end{figure}

Besides the clear absorption features mentioned in Table \ref{table:2}, \cite{tegle007} reported the detection of small bands located at 0.54, 0.58, and 0.6 $\mu$m. Our spectrum of Makemake only shows a hint of an absorption at 0.54 $\mu$m with a depth of about 2 \%, barely marginal over the point-to-point variation of the spectrum, {and therefore unreliable, while none of the other features could be detected (see Fig. \ref{figcomp})}.

\subsection{Wavelength shifts}\label{ws}

\begin{figure}
 \centering
 \includegraphics[width=\columnwidth]{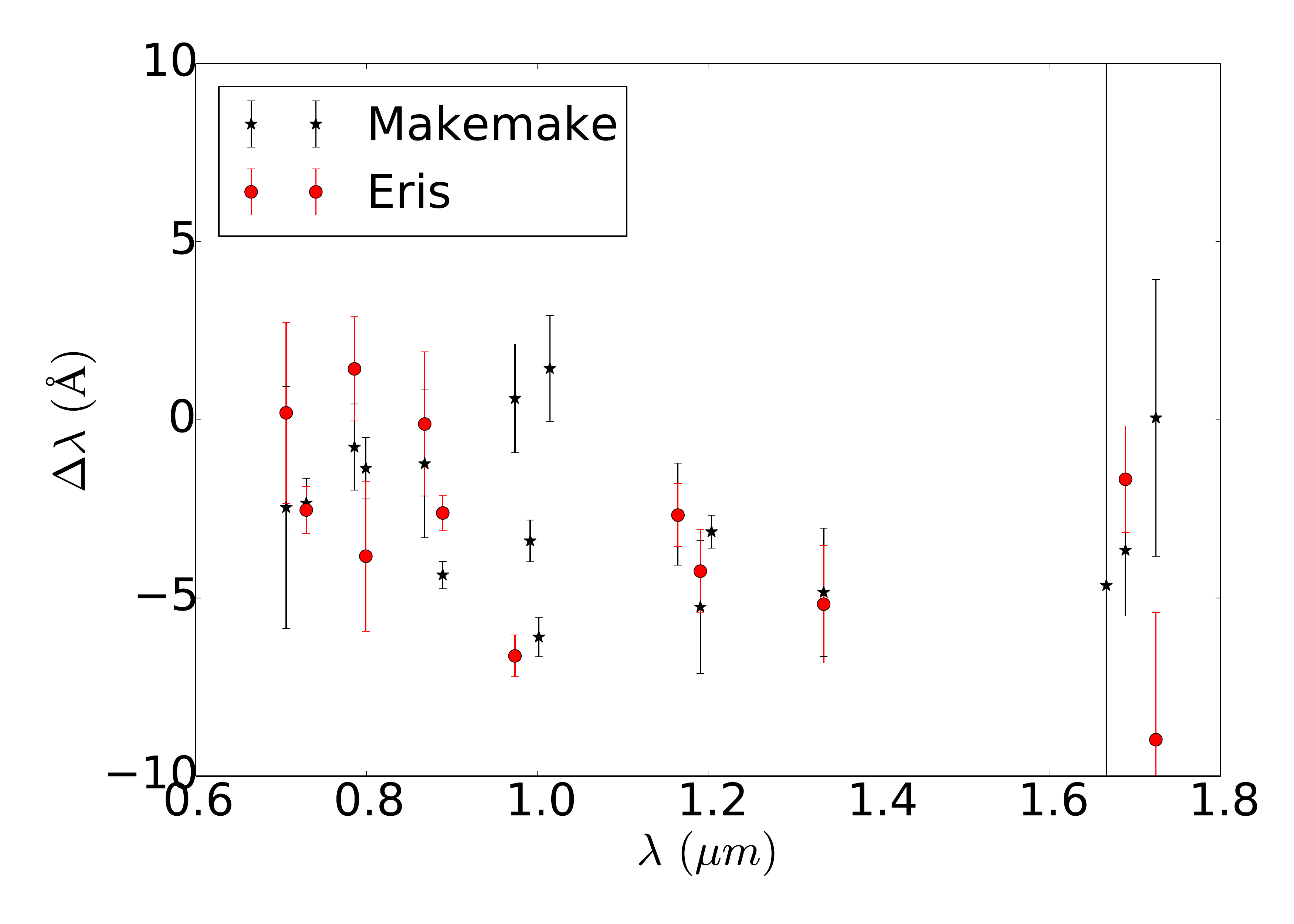}
 \caption{$\Delta\lambda$ vs. $\lambda$ for Makemake (black stars) and Eris (red circles).}\label{fig02}
\end{figure}

\begin{figure}
 \centering
 \includegraphics[width=\columnwidth]{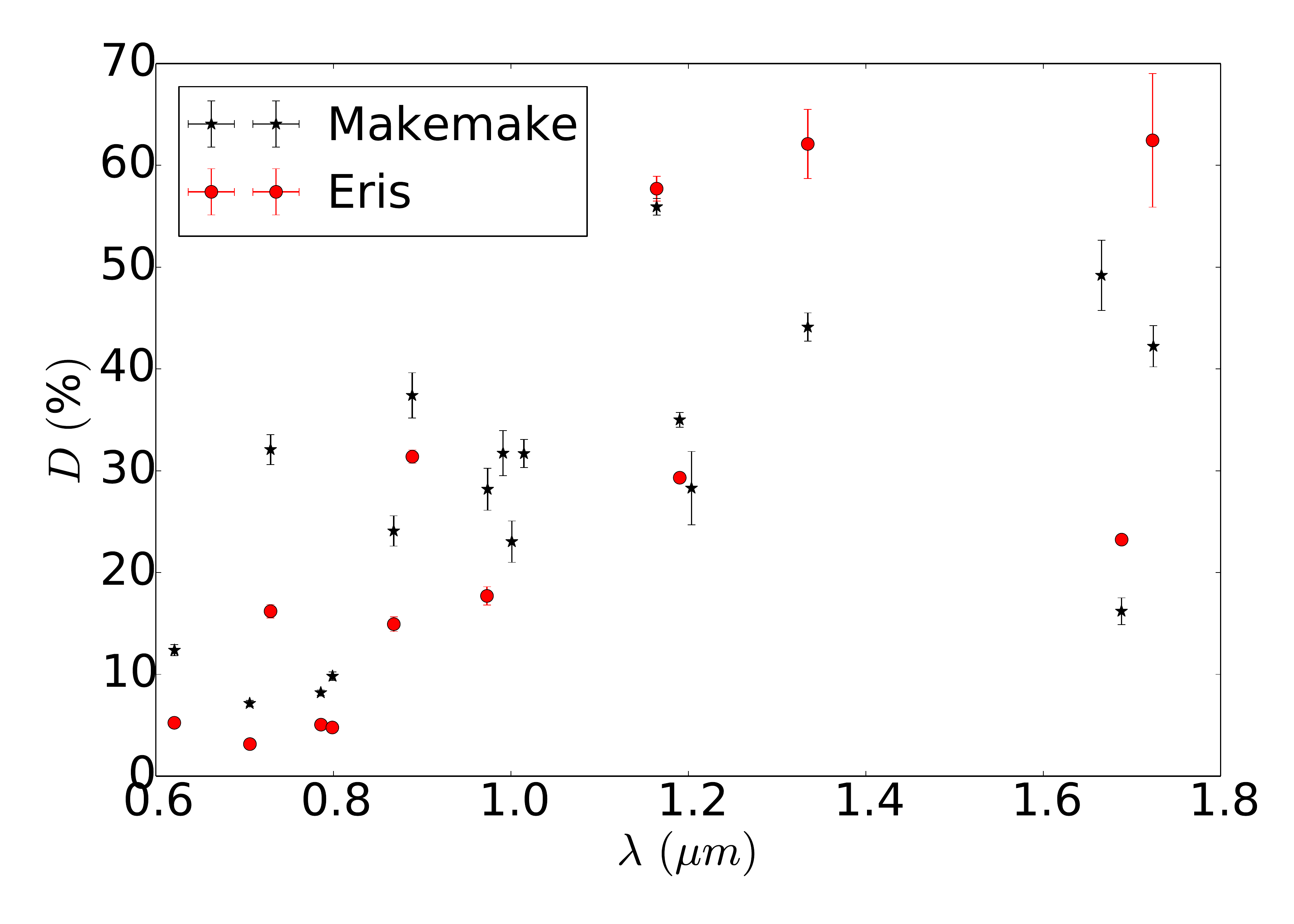}
 \caption{$D$ vs. $\lambda$ for Makemake (black stars) and Eris (red circles).}\label{fig05}
\end{figure}

We measured wavelength shifts ($\Delta\lambda$, expressed in \AA) and depths ($D$, expressed in \%) of the spectral features detected on the spectrum of Makemake by comparison with the spectrum of CH$_4$ ice obtained in laboratory. We analysed one by one several of the absorption features seen in Fig. \ref{fig01} with the following procedure: (i) We determine a linear continuum around the shoulders of the band of interest and divide the spectrum by this continuum. (ii) We select a small window around the apparent minimum of the feature and fit a second degree polynomial. The position of the minimum is taken as the zero of the first derivative of the polynomial. (iii) The process is repeated 10,000 times each time modifying the normalised flux values within a normal distribution with width equal to the nominal error at that wavelength. The final position is taken as the average value and its error of the flux as the standard deviation. (iv) To compute $D$ we used the weighed average flux, $f_w$, within the same window via
\begin{equation}\label{depth}
	D~[\%] = \big(1 - f_w^{-1}\big)\times100.
\end{equation}
(v) The positions of the CH$_4$ ice absorption bands are similarly measured. Note that, in this stage, we do not try to fit a specific model to the band of Makemake (or Eris). Instead, we use several models of CH$_4$ ice at 40 K \citep{grundy2002} for different grain sizes. We remove the local continuum in the same way as for our objects' data and define a small window around the minimum of the band. We fit a second degree polynomial within the window for all the models and choose the band position as the average and its error as the standard deviation. 
(vi) We obtain $$\Delta\lambda = \lambda_{\rm object} - \lambda_{\rm reference}.$$  
Values of $\Delta\lambda$ and $D$ for Makemake and Eris are shown in Table \ref{table:2} and displayed in Figs. \ref{fig02} and \ref{fig05} as function of wavelength.
\begin{table*}
\caption{Wavelength shifts and band depths. The first column shows the nominal central wavelength of the band, the second and third columns show the results for Makemake, the fourth and fifth for Eris, white the last columns shows the difference between Makemake and Eris. }\label{table:2}
\centering
\begin{tabular}{c c c c c c c}
\hline\hline
CH$_4$& Makemake & & Eris & &  M-E \\
$\lambda$ ($\mu$m)&$\Delta\lambda$ (\AA)&$D$ (\%)&$\Delta\lambda$ (\AA)&$D$ (\%)&$\Delta\lambda$ (\AA)\\
\hline
0.62&             & 12.4 (0.5)&            &  5.2 (0.2)& 1.7 (3.0)\\
0.71&  -2.5	(3.4) &  7.1 (0.3)&   0.2 (4.2)&  3.2 (0.1)&-2.7 (4.2)\\
0.73&  -2.3	(0.7) & 32.1 (1.5)&  -2.5 (1.0)& 16.2 (0.7)& 0.2 (1.0)\\
0.79&  -0.8	(1.2) &  8.2 (0.3)&   1.4 (1.9)&  5.1 (0.1)&-2.2 (1.9)\\
0.80&  -1.4	(0.9) &  9.8 (0.4)&  -3.8 (2.3)&  4.8 (0.2)& 2.5 (2.3)\\
0.87&  -1.2	(2.1) & 24.1 (1.5)&  -0.1 (2.9)& 14.9 (0.7)&-1.1 (2.9)\\
0.89&  -4.4	(0.4) & 37.4 (2.2)&  -2.6 (0.4)& 31.4 (0.7)&-1.7 (0.4)\\
0.97&   0.6	(1.5) & 28.2 (2.1)&  -6.6 (1.6)& 17.7 (0.9)& 7.2 (1.6)\\
0.99&  -3.4	(0.6) & 31.7 (2.2)&            &           &          \\
1.00&  -6.1	(0.6) & 23.0 (2.0)&            &           &          \\
1.01&   1.4	(1.5) & 31.7 (1.4)&            &           &          \\
1.16&  -2.6	(1.4) & 55.9 (0.8)&  -2.7 (1.7)& 57.7 (1.2)& 0.0 (1.7)\\
1.19&  -5.3	(1.9) & 35.0 (0.7)&  -4.2 (2.2)& 29.3 (0.5)&-1.0 (2.2)\\
1.20&  -3.1	(0.5) & 28.3 (3.6)&            &           &          \\
1.33&  -4.8	(1.8) & 44.1 (1.4)&  -5.2 (1.3)& 62.1 (3.4)& 0.3 (1.3)\\
1.67&  -4.6	(30.6)& 49.2 (3.5)&            & 62.7 (7.4)&          \\
1.69&  -3.7	(1.8) & 16.2 (1.3)&  -1.7 (1.4)& 23.2 (0.2)&-2.0 (1.4)\\
1.72&   0.1	(3.9) & 42.2 (2.0)&  -9.0 (1.6)& 62.5 (6.6)& 9.0 (1.6)\\
Average&-2.6 (2.1)&           &  -3.1 (2.8)&           & 0.8 (3.5)\\
\hline
\end{tabular}
\end{table*}

In Fig. \ref{fig02} it is apparent that Makemake follows a similar trend as already seen for Eris in \cite{alcan11}. We have not included in this work other comparisons, as done in our previous work, because our intent was to compare data of similar quality, wavelength coverage, and resolving power. 

\begin{figure}
 \centering
 \includegraphics[width=\columnwidth]{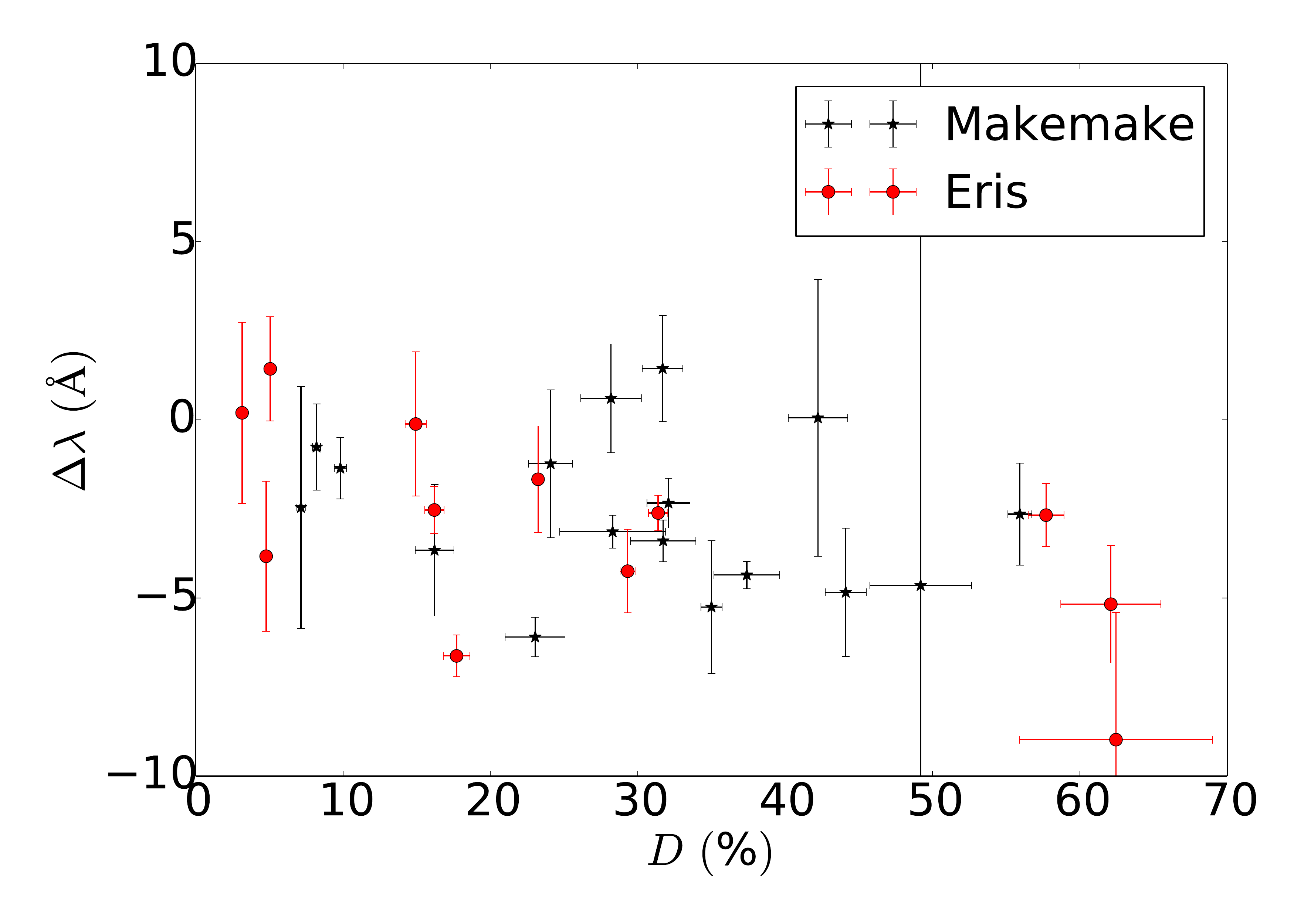}
 \caption{$\Delta\lambda$ vs. $D$ for Makemake (black stars) and Eris (red circles).}\label{fig03}
\end{figure}

Figure \ref{fig05} shows $D$ vs. $\lambda$.
Makemake tends to have deeper absorption bands than Eris, with the exception of the region beyond 1.4 $\mu$m where the  bands of Makemake are strongly saturated (except the 1.68 $\mu$m band) and, therefore, their depth becomes unreliable as, once saturated, the feature cannot grow deeper, instead its width increases. Deeper absorption features appear towards longer wavelength, which as pointed in many works \citep[for instance][and references therein]{alcan11} could be related to the thickness of the layer. Therefore, in Fig. \ref{fig03} we report the change of $\Delta\lambda$ with respect to $D$. It is possible to see that there is a small increase in the blue-shift with increasing depth, noticeable both on Makemake and Eris. To test this hypothesis we ran the Spearman test on both sets of data. The anti-correlation has a marginal statistically significance in the case of Eris, with a correlation parameter $r_s=-0.66$ and significance over $2\sigma$. No correlation is significant in the case of Makemake. 
\begin{figure}
 \centering
 \includegraphics[width=\columnwidth]{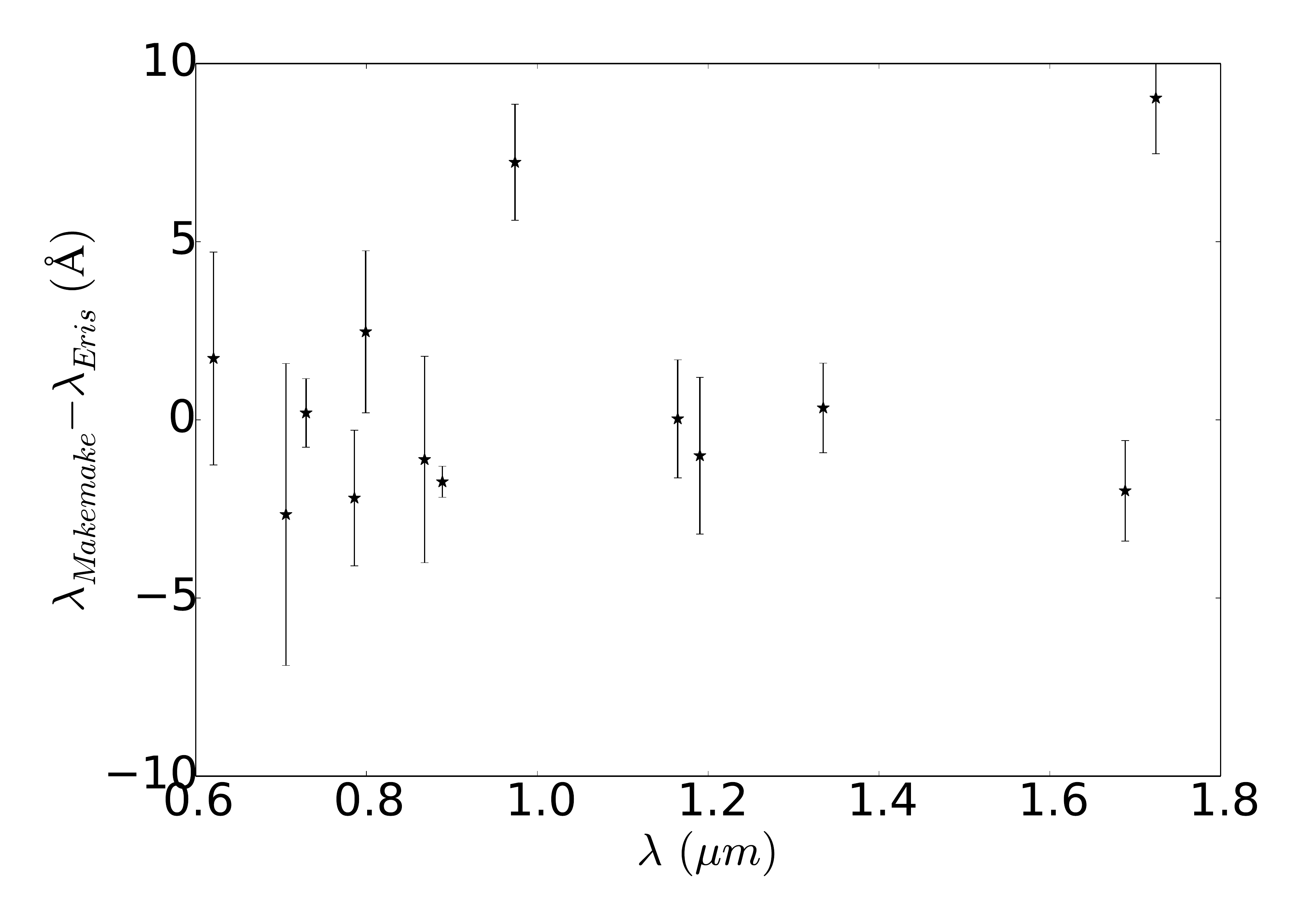}
 \caption{$\Delta\lambda$ vs. $\lambda$, comparison between Makemake and Eris.}\label{fig04}
\end{figure}
In the case of Eris we proposed that this could be indicative of a collapsed atmosphere onto its surface, could it be the same for Makemake? It seems difficult, because of its size and location which, in principle, preclude the existence of large quantities of volatile ices other than CH$_4$ on its surface.

Noteworthy is the remarkable fact that most of the $\Delta\lambda$ measured for Makemake and Eris are quite similar (Fig. \ref{fig04} and Table \ref{table:2}). The absorption features are mostly within three standard deviations from $\Delta\lambda=0$ \AA. We ran a 2 sample Kolmogorov-Smirnov test on the two sets, obtaining that we cannot reject the hypothesis that both $\Delta\lambda$-distributions come from the same parent distribution, with a significance over $2\sigma$. This, certainly, does not mean that the surfaces of Makemake and Eris are identical. 

\subsubsection{Cross-Correlation experiment}
\begin{figure}
 \centering
 \includegraphics[width=\columnwidth]{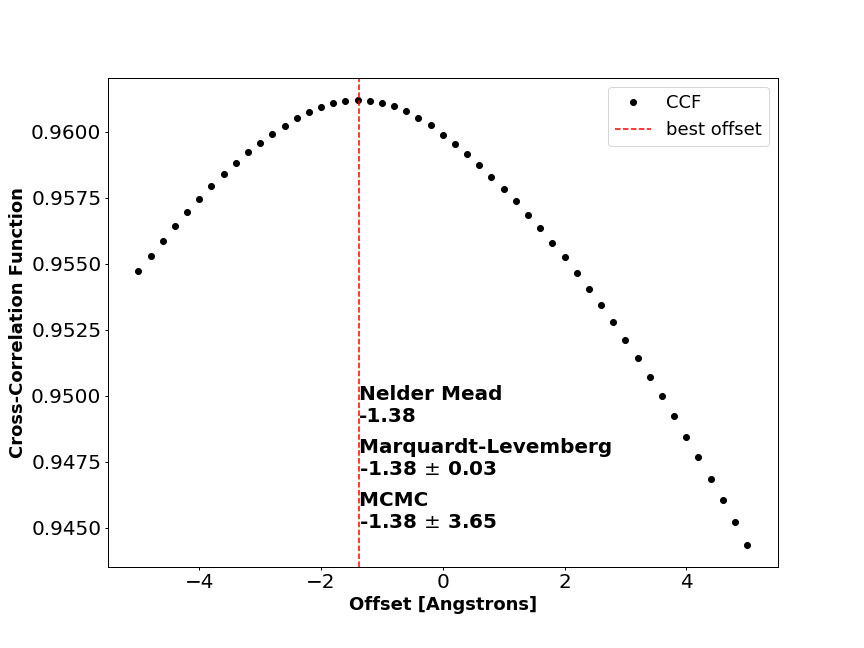}
 \caption{Makemake offset from Ice. }\label{fig_offset1}
\end{figure}

\begin{figure}
 \centering
 \includegraphics[width=\columnwidth]{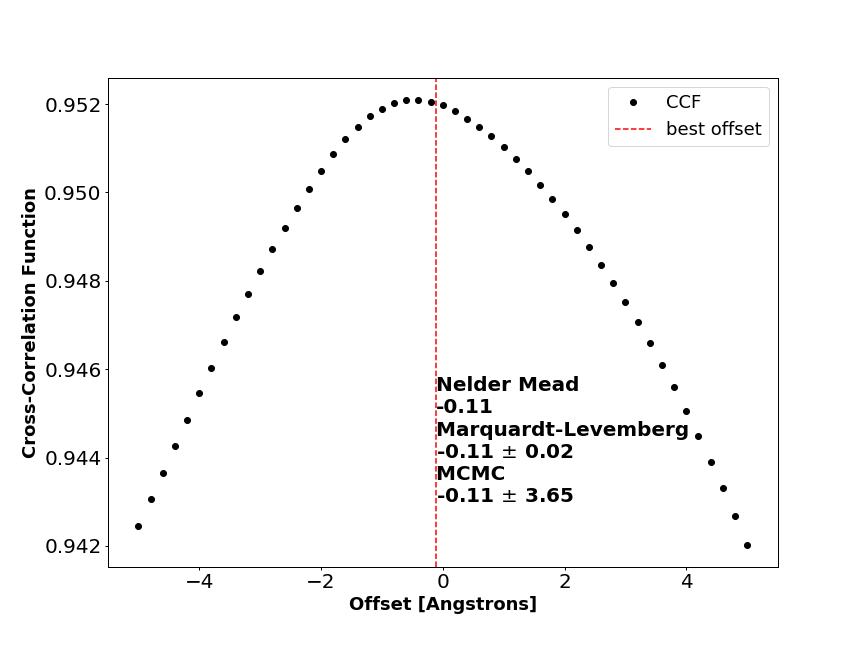}
 \caption{Makemake offset from Eris.}\label{fig_offset2}
\end{figure}

{
Aiming at double-checking our results from the previous section, we applied the cross-correlation function \citep{Brockwell2009}, CCF, to obtain the most likely shift between Makemake and CH$_4$ ice. We compute the errors of the shift from the confidence interval based on a Marquardt-Levemberg Algorithm \citep{Levenberg1944,Marquardt1963}. We also obtain a more conservative error from a Monte Carlo Markov Chain, MCMC, of $10,000$ nodes with $25$\% of acceptance \citep{emcee}. 

We applied shifts from $-25$ to $25$ steps in resolution bin of the spectra to obtain a measurement of the covariance between the Makemake spectra and the ice. We fit a second-degree polynomial to find the centre of the CCF, that corresponds to the maximum covariance and, therefore, to the most likely shift. We obtain the maximum CCF based on the Nelder-Mead Algorithm \citep{Nelder1965}. We use the shift from Nelder-Mead to obtain the confidence intervals based on Marquardt-Levemberg Non-Linear Least Squares algorithm and the MCMC.

The shift between Makemake and CH$_4$ is $-1.38 \pm 0.03$ \AA, with a more conservative result of $-1.38 \pm 3.38$ {\AA} (Fig. \ref{fig_offset1}).  We also applied the cross-correlation analysis between Makemake and Eris. The shift with Eris is $-0.11 \pm 0.02$ \AA, and assuming a conservative statistics, the shift is $-0.11 \pm 3.59$ {\AA} (Fig. \ref{fig_offset2}). The shift between Makemake and Eris is smaller than the variation in wavelength from each point in the resolution of Makemake spectra, which is $\Delta \lambda = 0.19$ \AA. In Table \ref{table:3} are summarised the results of the cross-correlation experiments.
These values are compatible with the averages shown in Table \ref{table:2}.

\begin{table}
\caption{Wavelength shifts from Cross-Validation Function between Makemake with our model, and Makemake with Eris.}\label{table:3}
\centering
\begin{tabular}{c c c}
\hline\hline
 Algorithm & $\Delta \lambda$ from Ice & $\Delta \lambda$ from Eris \\
 & \AA & \AA \\
\hline
Nelder-Mead & $-1.38$ & $-0.11$\\
Marquardt-Levemberg & $-1.38 \pm 0.03$ & $-0.11 \pm 0.02$\\
MCMC & $-1.38 \pm 3.38$ & $-0.11 \pm 3.59$\\
\hline
\end{tabular}
\end{table}

}

\subsection{Spectral Modelling}
\begin{figure}
    \centering
    \includegraphics[width=\columnwidth]{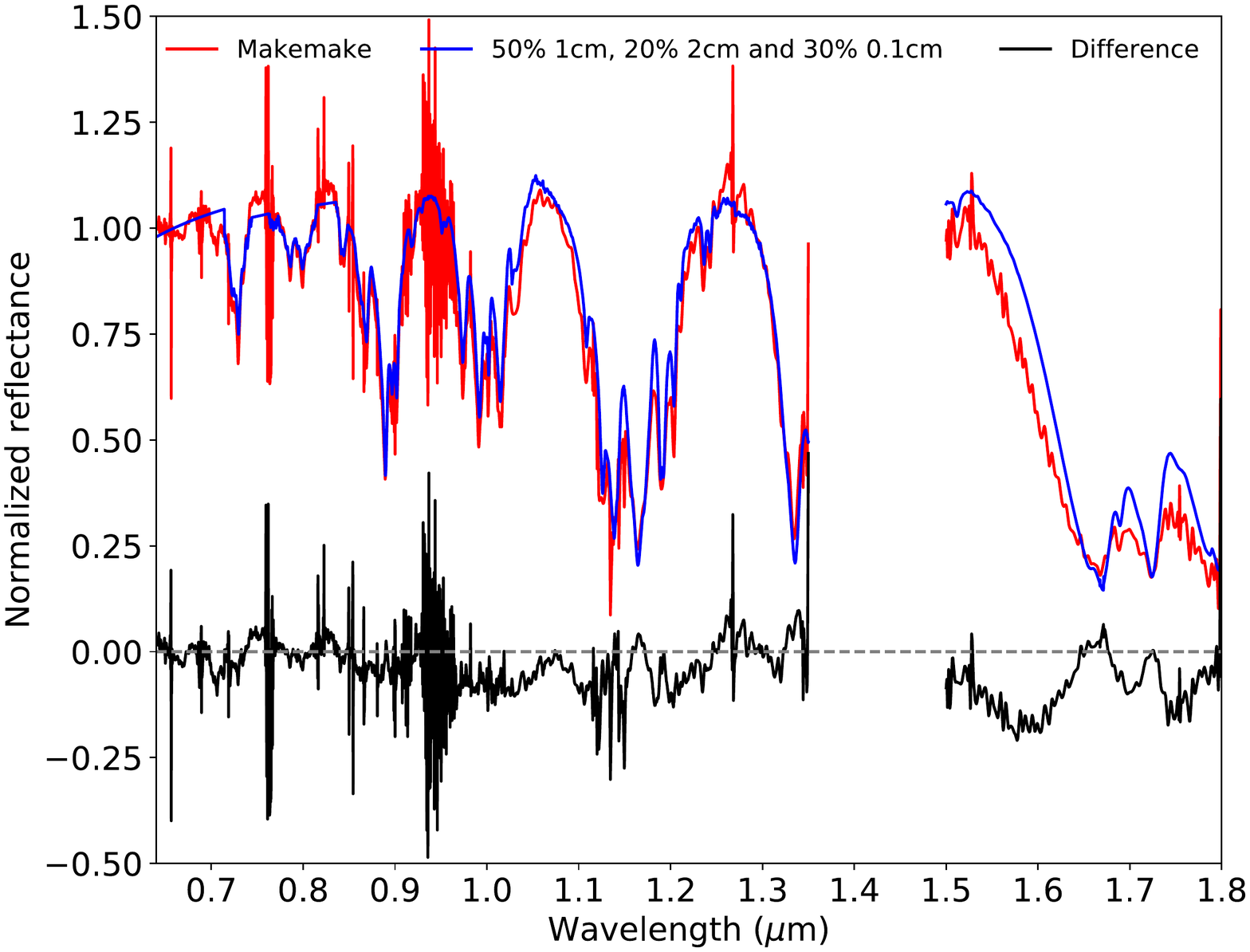}
    \caption{{Normalised spectrum of Makemake (red) compared with a synthetic spectra (blue) made with different proportions and size grains of CH$_4$ (see the text). The difference between the original and synthetic spectra is shown in black. The region close to 1.4 $\mu$m was masked due to the absorption of the atmosphere.}}\label{makemake_mod}
\end{figure}

\begin{figure}
    \centering
    \includegraphics[width=\columnwidth]{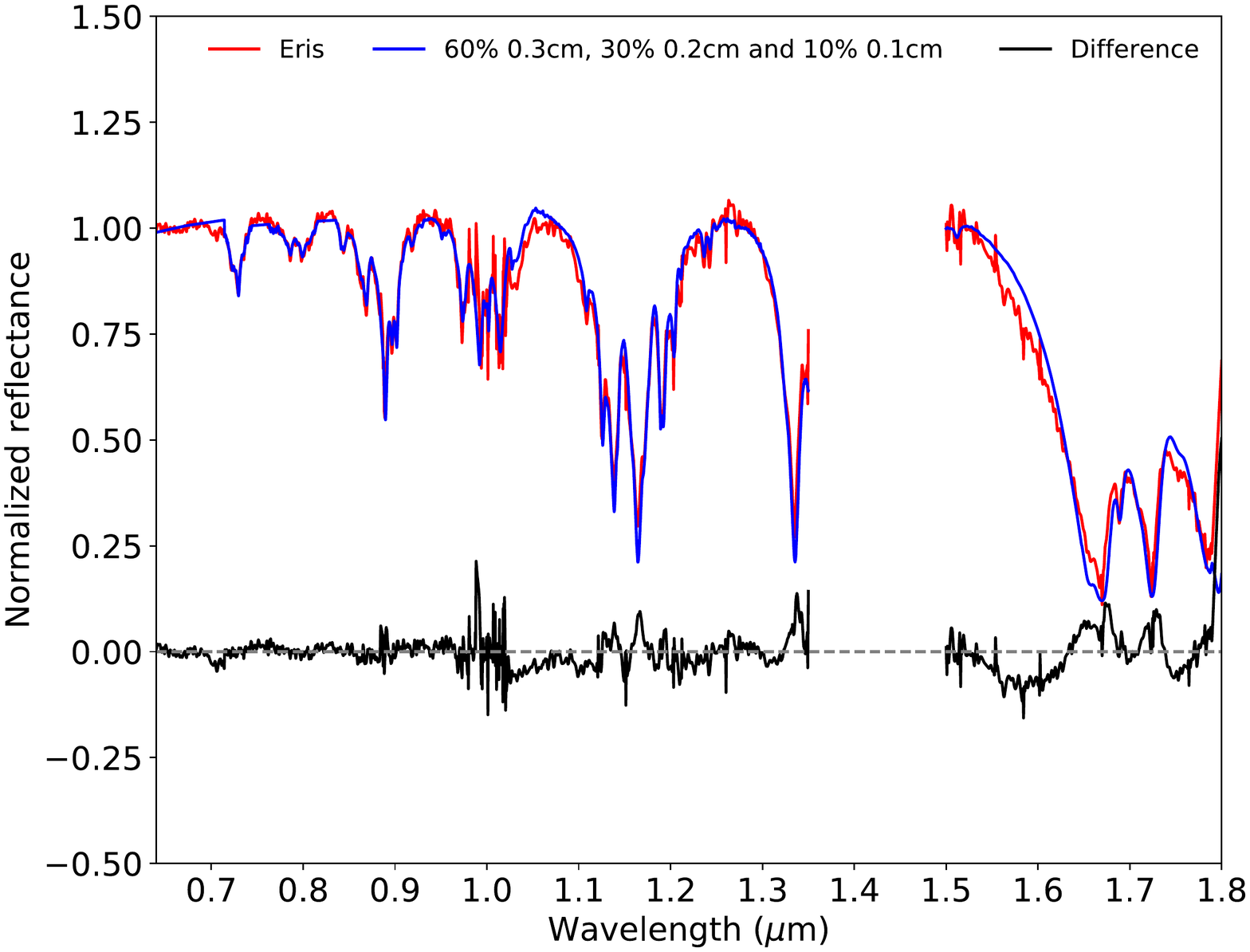}
    \caption{{Normalised spectrum of Eris (red) compared with a synthetic spectra (blue) made with different proportions and size grains of CH$_4$ (see the text). The difference between the original and synthetic spectra is shown in black. The region close to 1.4 $\mu$m was masked due to the absorption of the atmosphere.}}\label{eris_mod}
\end{figure}

We created synthetic spectra of Makemake and Eris 
{{using the Hapke radiative transfer model \citep{hapke93} with optical constants of pure CH$_4$ ice at 40 K \citep{grundy2002}. Other scattering theories exist \cite[e.g.,][]{doute98,shkuratov99} that could provide similar quality of fits to the data as those presented here, but with different percentages and grain sizes of the components \citep{poulet02}. However, we decided to interpret our data with a simple model since it provides a reasonable fit to both spectra and we are not aiming at a detailed description of their surface composition, which has been extensively studied elsewhere (see references in the Sect. \ref{sec:introduction}). Because our goal is to perform a comparative study between Makemake and Eris we only use different combinations of pure CH$_4$ ice.} Also, the synthetic spectra are limited to $\lambda>0.7$ $\mu$m.

For the purpose of this work, we assumed that CH$_4$ ice is spatially segregated on the surfaces of Makemake and Eris. The model that best fits the spectrum of Makemake contains 50\% of CH$_4$ ice of 1 cm grains, 30\% of 2 cm grains, and 20\% with grains of 0.1 cm (Fig. \ref{makemake_mod}). The model provides a close description of the data. However, there are some remaining differences, especially in the 1.5 to 1.7 $\mu$m range. The spectral model of Eris (Fig. \ref{eris_mod}) contains 60\% of CH$_4$ ice with grains of 0.3 cm, 30\% with 0.2 cm grains, and 10\% with grains of 0.1 cm. With exception of the spectral range between 1.5 to 1.7 $\mu$m, where apparently an extra absorption is needed, the model describes the properties of the reflectance of the surface satisfactorily.} 

\subsection{NIR spectrum interpretation}
The optical properties of the spectra of Makemake and Eris are dominated by CH$_4$ ice. In the case of Eris, its radius \cite[$1,163 \pm 6$ km,][]{sicardy2011} and temperature about 35 K \citep{sicardy2011} support the retention of volatile ices \citep{schal07}, however no direct detection other than CH$_4$ has been possible to date. In particular N$_2$ or CO ices might be found, but the low temperature keeps these ices in their $\alpha$-state, whose spectral features cannot be resolved by X-Shooter \citep[see][]{alcan11}.
On the other hand, considering Makemake's size ($\sim1400$ km) and surface temperature [36 K, if it were a slow rotator \cite[see][supplemental information]{ortiz12}], the retention regime for this object is different than the retention regime for Eris \citep{brown2012AREPS}. Makemake is capable of retaining CH$_4$ ice but is not expected to retain large amounts of N$_2$ or CO ices, so their direct detection in its near-infrared spectrum is unlikely \citep{brown07}. 

The molecule of CH$_4$ ice is optically very active, therefore, the presence of other components are masked by its absorption bands. Nevertheless, Eris' and Makemake's CH$_4$ bands are seen to be partially shifted to shorter wavelengths relative to the wavelengths of pure CH$_4$ ice absorption bands (see Sect.\ref{ws}), {indicating that CH$_4$ and N$_2$ are present on Eris and Makemake.}

\section{Discussion}
{Our spectrum of Makemake does not confirm the detection of the subtle absorption bands of CH$_4$ ice short-ward of 0.62 $\mu$m proposed by \cite{tegle007}. One possible explanation is the lower SNR of our spectrum, although with a larger resolving power.}
Nevertheless, it must be kept in mind that these bands are expected to exist \citep[e.g.,][]{patel1980}, but they would be extremely hard to detect due to the large path lengths necessary to produce them.

As mentioned in the Introduction, neither \cite{loren15} nor \cite{perna17} detected any significant heterogeneity on the surface of Makemake. Both works used rotational resolved spectroscopy with good resolving power in the visible part. Unfortunately, we do not have rotational resolved spectra of Makemake, but we do cover a wider spectral range with higher resolving power. Our data show that the centres of the absorption features of CH$_4$ ice seen on the spectra of Makemake and Eris have remarkably similar blue-shifts, when compared with CH$_4$ ice measured in laboratory. Furthermore, {Makemake's band centres are marginally bluer than Eris', which seemed {\it a priori} unlikely because Eris should have retained more volatile ices than Makemake due to its size and location, {and it is unexpected \cite[e.g.][pg 130]{Young2020}}}. Nevertheless, if the N$_2$ is in different phases, that might explain, at least partially these shifts. In \cite{quiri97}, the shifts of the features of CH$_4$ ice are slightly larger if the N$_2$ ice is in its $\alpha$-phase, as expected for Eris. Interestingly, Makemake follows a similar trend in the $\Delta\lambda$ vs. $D$ space as Eris: deeper adsorptions show larger shifts, although with less statistical significance. We do not believe this is pointing to a collapsed frozen atmosphere on the surface of Makemake, but rather to the scatter of our data.

The spectral slope in the visible range of Makemake is larger than the Eris, suggesting that more processed material is in fact present on the surface of Makemake. The rough spectral modelling performed on both spectra show large residuals (Figs. \ref{makemake_mod} and \ref{eris_mod}), especially in the $1.5-1.7$ $\mu$m. We attribute these differences to minor components on the surface of these objects, as proposed by \cite{brown15}. They showed that high-order hydrocarbons are present in the spectrum of Makemake, in particular ethane (C$_2$H$_6$), that improve the modelling of the spectrum \citep[see][]{perna17}.

\section{Conclusions}
In this work we present X-Shooter data of the dwarf planet Makemake which, to the best of our knowledge, has the greater resolving power over a large spectral range. The spectrum is compared with that of CH$_4$ ice and that of Eris, obtained with the same instrument and similar observational setup.

The modelling of the spectrum shows the need for more ingredients besides pure CH$_4$ ice, not only for the large residuals, but also for the location of the absorption bands.
{Interestingly, we see that the location of the features in Makemake and Eris are remarkably similar. Furthermore, Makemake's are slightly blue-shifted, with respect to Eris', instead of red-shifted, as it was usually expected. This could only be achieved with mid-resolving power spectroscopy. } 

{The wavelength shifts could be affected by the reservoir of volatile on the surface of both dwarf-planets and by the temperature of their surfaces over their orbits, especially in the passages by their perihelia. Furthermore, the temperature is key in the phase of the N$_2$ ice, which experiences the transition from $\alpha$ to $\beta$ phase at 35.6 K. The low-temperature $\alpha$-phase ice has much deeper and narrower absorption bands and is likely to be the dominant phase of N$_2$ at Eris \citep{alcan11}, whereas, Makemake's with higher surface temperature, and closer to the Sun, is more likely to contain the $\beta$-phase of N$_2$ ice. The fact that the shifts of the bands, for both Makemake and Eris, are larger for the deepest bands is an indicative that the mixture of CH$_4$ and N$_2$ must be more abundant in the subsurface layers so, while the surface of Eris could be covered by a richer N$_2$ layer, product of the collapse of an atmosphere, Makemake, because of its redder colour, would be richer in the products of the irradiation of CH$_4$, the tholins. What is the actual nature of these tholins is something that JWST will be able to investigate, same as for the presence of the lower temperature N2-ice, never directly detected in the solar system before.}

\section*{Acknowledgements}
\noindent
We thank the thorough review made by an anonymous referee that helped to improve this work.\\
\noindent
Facilities: Based on observations collected at the European Organisation for Astronomical Research in the Southern Hemisphere under ESO programme 091.C-0381(A), P.I. AAC.\\
\noindent
Funding:
AAC acknowledges support from FAPERJ (grant E26/203.186/2016), CNPq (grants  304971/2016-2  and 401669/2016-5), and the Universidad de Alicante (contract UATALENTO18-02).
ACSF and WMF acknowledge support from CAPES.
NPA acknowledges support from SRI/FSI funds through the project ``Digging-Up Ice Rocks in the Solar System''.
JLO thanks support from grant AYA2017-89637-R and from the State Agency for Research of the Spanish MCIU through the ``Center of Excellence Severo Ochoa'' award for the Instituto de Astrof\'isica de Andaluc\'ia (SEV-2017-0709).\\
\noindent
Software:
IRAF is distributed by the National Optical Astronomy Observatory, which is operated by the Association of Universities for Research in Astronomy (AURA) under a cooperative agreement with the National Science Foundation.
https://www.python.org/.
https://www.scipy.org/.
Matplotlib \citep{hunte2007}. 

\section*{Data Availability}
The data underlying this article will be shared on reasonable request to the corresponding author.




\bibliographystyle{mnras}
\bibliography{makemake} 

\begin{thebibliography}{}
\makeatletter
\relax
\def\mn@urlcharsother{\let\do\@makeother \do\$\do\&\do\#\do\^\do\_\do\%\do\~}
\def\mn@doi{\begingroup\mn@urlcharsother \@ifnextchar [ {\mn@doi@}
  {\mn@doi@[]}}
\def\mn@doi@[#1]#2{\def\@tempa{#1}\ifx\@tempa\@empty \href
  {http://dx.doi.org/#2} {doi:#2}\else \href {http://dx.doi.org/#2} {#1}\fi
  \endgroup}
\def\mn@eprint#1#2{\mn@eprint@#1:#2::\@nil}
\def\mn@eprint@arXiv#1{\href {http://arxiv.org/abs/#1} {{\tt arXiv:#1}}}
\def\mn@eprint@dblp#1{\href {http://dblp.uni-trier.de/rec/bibtex/#1.xml}
  {dblp:#1}}
\def\mn@eprint@#1:#2:#3:#4\@nil{\def\@tempa {#1}\def\@tempb {#2}\def\@tempc
  {#3}\ifx \@tempc \@empty \let \@tempc \@tempb \let \@tempb \@tempa \fi \ifx
  \@tempb \@empty \def\@tempb {arXiv}\fi \@ifundefined
  {mn@eprint@\@tempb}{\@tempb:\@tempc}{\expandafter \expandafter \csname
  mn@eprint@\@tempb\endcsname \expandafter{\@tempc}}}

\bibitem[\protect\citeauthoryear{{Alvarez-Candal} et~al.,}{{Alvarez-Candal}
  et~al.}{2011}]{alcan11}
{Alvarez-Candal} A.,  et~al., 2011, \mn@doi [\aap]
  {10.1051/0004-6361/201117069}, \href
  {http://adsabs.harvard.edu/abs/2011A%26A...532A.130A} {532, A130}

\bibitem[\protect\citeauthoryear{{Bierson} \& {Nimmo}}{{Bierson} \&
  {Nimmo}}{2019}]{bierson2019}
{Bierson} C.~J.,  {Nimmo} F.,  2019, \mn@doi [\icarus]
  {10.1016/j.icarus.2019.01.027}, \href
  {https://ui.adsabs.harvard.edu/abs/2019Icar..326...10B} {326, 10}

\bibitem[\protect\citeauthoryear{Brockwell \& Davis}{Brockwell \&
  Davis}{2009}]{Brockwell2009}
Brockwell P.~J.,  Davis R.~A.,  2009, Time Series: Theory and Methods (Springer
  Series in Statistics).
Springer

\bibitem[\protect\citeauthoryear{{Brown}}{{Brown}}{2012}]{brown2012AREPS}
{Brown} M.~E.,  2012, \mn@doi [Annual Review of Earth and Planetary Sciences]
  {10.1146/annurev-earth-042711-105352}, \href
  {https://ui.adsabs.harvard.edu/abs/2012AREPS..40..467B} {40, 467}

\bibitem[\protect\citeauthoryear{Brown}{Brown}{2013}]{brown13}
Brown M.~E.,  2013, \mn@doi [The Astrophysical Journal]
  {10.1088/2041-8205/767/1/L7}, 767, L7

\bibitem[\protect\citeauthoryear{Brown, Barkume, Blake, Schaller, Rabinowitz,
  Roe  \& Trujillo}{Brown et~al.}{2007}]{brown07}
Brown M.~E.,  Barkume K.~M.,  Blake G.~A.,  Schaller E.~L.,  Rabinowitz D.~L.,
  Roe H.~G.,   Trujillo C.~A.,  2007, \mn@doi [The Astronomical Journal]
  {10.1086/509734}, 133, 284

\bibitem[\protect\citeauthoryear{Brown, Schaller  \& Blake}{Brown
  et~al.}{2015}]{brown15}
Brown M.~E.,  Schaller E.~L.,   Blake G.~A.,  2015, \mn@doi [The Astronomical
  Journal] {10.1088/0004-6256/149/3/105}, 149, 105

\bibitem[\protect\citeauthoryear{{Brunetto}, {Caniglia}, {Baratta}  \&
  {Palumbo}}{{Brunetto} et~al.}{2008}]{brunetto2008}
{Brunetto} R.,  {Caniglia} G.,  {Baratta} G.~A.,   {Palumbo} M.~E.,  2008,
  \mn@doi [\apj] {10.1086/591509}, \href
  {https://ui.adsabs.harvard.edu/abs/2008ApJ...686.1480B} {686, 1480}

\bibitem[\protect\citeauthoryear{{De Pra}, {Carvano}, {Morate}, {Licandro}  \&
  {Pinilla-Alonso}}{{De Pra} et~al.}{2018}]{cana}
{De Pra} M.,  {Carvano} J.,  {Morate} D.,  {Licandro} J.,   {Pinilla-Alonso}
  N.,  2018.

\bibitem[\protect\citeauthoryear{{Dout{\'e}} \& {Schmitt}}{{Dout{\'e}} \&
  {Schmitt}}{1998}]{doute98}
{Dout{\'e}} S.,  {Schmitt} B.,  1998, \mn@doi [\jgr] {10.1029/98JE01894}, \href
  {https://ui.adsabs.harvard.edu/abs/1998JGR...10331367D} {103, 31367}

\bibitem[\protect\citeauthoryear{Eluszkiewicz, Cady-Pereira, Brown  \&
  Stansberry}{Eluszkiewicz et~al.}{2007}]{elusz07}
Eluszkiewicz J.,  Cady-Pereira K.,  Brown M.~E.,   Stansberry J.~A.,  2007,
  \mn@doi [Journal of Geophysical Research] {10.1029/2007JE002892}, 112, E06003

\bibitem[\protect\citeauthoryear{{Foreman-Mackey}, {Hogg}, {Lang}  \&
  {Goodman}}{{Foreman-Mackey} et~al.}{2013}]{emcee}
{Foreman-Mackey} D.,  {Hogg} D.~W.,  {Lang} D.,   {Goodman} J.,  2013, \mn@doi
  [\pasp] {10.1086/670067}, \href
  {http://adsabs.harvard.edu/abs/2013PASP..125..306F} {125, 306}

\bibitem[\protect\citeauthoryear{{Grundy}, {Schmitt}  \& {Quirico}}{{Grundy}
  et~al.}{2002}]{grundy2002}
{Grundy} W.~M.,  {Schmitt} B.,   {Quirico} E.,  2002, \mn@doi [\icarus]
  {10.1006/icar.2001.6726}, \href
  {https://ui.adsabs.harvard.edu/abs/2002Icar..155..486G} {155, 486}

\bibitem[\protect\citeauthoryear{{Hainaut}, {Boehnhardt}  \&
  {Protopapa}}{{Hainaut} et~al.}{2012}]{mboss}
{Hainaut} O.~R.,  {Boehnhardt} H.,   {Protopapa} S.,  2012, \mn@doi [\aap]
  {10.1051/0004-6361/201219566}, \href
  {https://ui.adsabs.harvard.edu/abs/2012A&A...546A.115H} {546, A115}

\bibitem[\protect\citeauthoryear{{Hapke}}{{Hapke}}{1993}]{hapke93}
{Hapke} B.,  1993, {Theory of reflectance and emittance spectroscopy}

\bibitem[\protect\citeauthoryear{Heinze \& DeLahunta}{Heinze \&
  DeLahunta}{2009}]{heinz09}
Heinze A.~N.,  DeLahunta D.,  2009, \mn@doi [The Astronomical Journal]
  {10.1088/0004-6256/138/2/428}, 138, 428

\bibitem[\protect\citeauthoryear{{Hromakina} et~al.,}{{Hromakina}
  et~al.}{2019}]{hromakina2019}
{Hromakina} T.~A.,  et~al., 2019, \mn@doi [\aap] {10.1051/0004-6361/201935274},
  \href {https://ui.adsabs.harvard.edu/abs/2019A&A...625A..46H} {625, A46}

\bibitem[\protect\citeauthoryear{Hunter}{Hunter}{2007}]{hunte2007}
Hunter J.~D.,  2007, \mn@doi [Computing In Science \& Engineering]
  {10.1109/MCSE.2007.55}, 9, 90

\bibitem[\protect\citeauthoryear{Khare, Sagan, Arakawa, Suits, Callcott  \&
  Williams}{Khare et~al.}{1984}]{khare84}
Khare B.,  Sagan C.,  Arakawa E.,  Suits F.,  Callcott T.,   Williams M.,
  1984, Icarus, 60, 127

\bibitem[\protect\citeauthoryear{Levenberg}{Levenberg}{1944}]{Levenberg1944}
Levenberg K.,  1944, \mn@doi [Quarterly of Applied Mathematics]
  {10.1090/qam/10666}, 2, 164

\bibitem[\protect\citeauthoryear{Licandro, Pinilla-Alonso, Pedani, Oliva, Tozzi
   \& Grundy}{Licandro et~al.}{2006}]{lican06}
Licandro J.,  Pinilla-Alonso N.,  Pedani M.,  Oliva E.,  Tozzi G.~P.,   Grundy
  W.~M.,  2006, \mn@doi [Astronomy {\&} Astrophysics]
  {10.1051/0004-6361:200500219}, 445, L35

\bibitem[\protect\citeauthoryear{Lorenzi, Pinilla-Alonso  \& Licandro}{Lorenzi
  et~al.}{2015}]{loren15}
Lorenzi V.,  Pinilla-Alonso N.,   Licandro J.,  2015, \mn@doi [Astronomy {\&}
  Astrophysics] {10.1051/0004-6361/201425575}, 577, A86

\bibitem[\protect\citeauthoryear{{Lorenzi}, {Pinilla-Alonso}, {Licandro},
  {Cruikshank}, {Grundy}, {Binzel}  \& {Emery}}{{Lorenzi}
  et~al.}{2016}]{lorenzi2016}
{Lorenzi} V.,  {Pinilla-Alonso} N.,  {Licandro} J.,  {Cruikshank} D.~P.,
  {Grundy} W.~M.,  {Binzel} R.~P.,   {Emery} J.~P.,  2016, \mn@doi [\aap]
  {10.1051/0004-6361/201527281}, \href
  {https://ui.adsabs.harvard.edu/abs/2016A&A...585A.131L} {585, A131}

\bibitem[\protect\citeauthoryear{Marquardt}{Marquardt}{1963}]{Marquardt1963}
Marquardt D.~W.,  1963, \mn@doi [Journal of the Society for Industrial and
  Applied Mathematics] {10.1137/0111030}, 11, 431

\bibitem[\protect\citeauthoryear{{Merlin}, {Barucci}, {de Bergh}, {DeMeo},
  {Alvarez-Candal}, {Dumas}  \& {Cruikshank}}{{Merlin} et~al.}{2010}]{merli10}
{Merlin} F.,  {Barucci} M.~A.,  {de Bergh} C.,  {DeMeo} F.~E.,
  {Alvarez-Candal} A.,  {Dumas} C.,   {Cruikshank} D.~P.,  2010, \mn@doi
  [\icarus] {10.1016/j.icarus.2010.07.028}, \href
  {http://adsabs.harvard.edu/abs/2010Icar..210..930M} {210, 930}

\bibitem[\protect\citeauthoryear{Nelder \& Mead}{Nelder \&
  Mead}{1965}]{Nelder1965}
Nelder J.~A.,  Mead R.,  1965, \mn@doi [The Computer Journal]
  {10.1093/comjnl/7.4.308}, 7, 308

\bibitem[\protect\citeauthoryear{Ortiz et~al.,}{Ortiz et~al.}{2012}]{ortiz12}
Ortiz J.~L.,  et~al., 2012, \mn@doi [Nature] {10.1038/nature11597}, 491, 566

\bibitem[\protect\citeauthoryear{Parker, Buie, Grundy  \& Noll}{Parker
  et~al.}{2016}]{parke16}
Parker A.~H.,  Buie M.~W.,  Grundy W.~M.,   Noll K.~S.,  2016, \mn@doi [The
  Astrophysical Journal] {10.3847/2041-8205/825/1/L9}, 825, L9

\bibitem[\protect\citeauthoryear{{Patel}, {Nelson}  \& {Kerl}}{{Patel}
  et~al.}{1980}]{patel1980}
{Patel} C.~K.~N.,  {Nelson} E.~T.,   {Kerl} R.~J.,  1980, \mn@doi [\nat]
  {10.1038/286368a0}, \href
  {https://ui.adsabs.harvard.edu/abs/1980Natur.286..368P} {286, 368}

\bibitem[\protect\citeauthoryear{Perna, Hromakina, Merlin, Ieva, Fornasier,
  Belskaya  \& {Mazzotta Epifani}}{Perna et~al.}{2017}]{perna17}
Perna D.,  Hromakina T.,  Merlin F.,  Ieva S.,  Fornasier S.,  Belskaya I.,
  {Mazzotta Epifani} E.,  2017, \mn@doi [Monthly Notices of the Royal
  Astronomical Society] {10.1093/mnras/stw3272}, 466, 3594

\bibitem[\protect\citeauthoryear{Poulet, Cuzzi, Cruikshank, Roush  \&
  Dalle~Ore}{Poulet et~al.}{2002}]{poulet02}
Poulet F.,  Cuzzi J.,  Cruikshank D.,  Roush T.,   Dalle~Ore C.,  2002, Icarus,
  160, 313

\bibitem[\protect\citeauthoryear{{Protopapa}, {Grundy}, {Tegler}  \&
  {Bergonio}}{{Protopapa} et~al.}{2015}]{protopapa2015}
{Protopapa} S.,  {Grundy} W.~M.,  {Tegler} S.~C.,   {Bergonio} J.~M.,  2015,
  \mn@doi [\icarus] {10.1016/j.icarus.2015.02.027}, \href
  {https://ui.adsabs.harvard.edu/abs/2015Icar..253..179P} {253, 179}

\bibitem[\protect\citeauthoryear{{Quirico} \& {Schmitt}}{{Quirico} \&
  {Schmitt}}{1997}]{quiri97}
{Quirico} E.,  {Schmitt} B.,  1997, \mn@doi [\icarus] {10.1006/icar.1996.5663},
  \href {http://adsabs.harvard.edu/abs/1997Icar..127..354Q} {127, 354}

\bibitem[\protect\citeauthoryear{{Ram{\'\i}rez} et~al.,}{{Ram{\'\i}rez}
  et~al.}{2012}]{ramirez2012}
{Ram{\'\i}rez} I.,  et~al., 2012, \mn@doi [\apj] {10.1088/0004-637X/752/1/5},
  \href {https://ui.adsabs.harvard.edu/abs/2012ApJ...752....5R} {752, 5}

\bibitem[\protect\citeauthoryear{Schaller \& Brown}{Schaller \&
  Brown}{2007}]{schal07}
Schaller E.~L.,  Brown M.~E.,  2007, \mn@doi [The Astrophysical Journal]
  {10.1086/516709}, 659, L61

\bibitem[\protect\citeauthoryear{Shkuratov, Starukhina, Hoffmann  \&
  Arnold}{Shkuratov et~al.}{1999}]{shkuratov99}
Shkuratov Y.,  Starukhina L.,  Hoffmann H.,   Arnold G.,  1999, Icarus, 137,
  235

\bibitem[\protect\citeauthoryear{{Sicardy} et~al.,}{{Sicardy}
  et~al.}{2011}]{sicardy2011}
{Sicardy} B.,  et~al., 2011, \mn@doi [\nat] {10.1038/nature10550}, \href
  {https://ui.adsabs.harvard.edu/abs/2011Natur.478..493S} {478, 493}

\bibitem[\protect\citeauthoryear{Simonia \& Cruikshank}{Simonia \&
  Cruikshank}{2018}]{simonia2018}
Simonia I.,  Cruikshank D.~P.,  2018, Open Astronomy, 27, 341

\bibitem[\protect\citeauthoryear{{Souza-Feliciano}, {Alvarez-Candal}  \&
  {Jim{\'e}nez-Teja}}{{Souza-Feliciano} et~al.}{2018}]{souza18}
{Souza-Feliciano} A.~C.,  {Alvarez-Candal} A.,   {Jim{\'e}nez-Teja} Y.,  2018,
  \mn@doi [\aap] {10.1051/0004-6361/201731464}, \href
  {https://ui.adsabs.harvard.edu/abs/2018A&A...614A..92S} {614, A92}

\bibitem[\protect\citeauthoryear{{Starck} \& {Murtagh}}{{Starck} \&
  {Murtagh}}{2006}]{starc06}
{Starck} J.-L.,  {Murtagh} F.,  2006, {Astronomical Image and Data Analysis},
  \mn@doi{10.1007/978-3-540-33025-7.
}

\bibitem[\protect\citeauthoryear{{Tan} \& {Kargel}}{{Tan} \&
  {Kargel}}{2018}]{tan2018}
{Tan} S.~P.,  {Kargel} J.~S.,  2018, \mn@doi [\mnras] {10.1093/mnras/stx3036},
  \href {https://ui.adsabs.harvard.edu/abs/2018MNRAS.474.4254T} {474, 4254}

\bibitem[\protect\citeauthoryear{Tegler, Grundy, Romanishin, Consolmagno,
  Mogren  \& Vilas}{Tegler et~al.}{2007}]{tegle007}
Tegler S.~C.,  Grundy W.~M.,  Romanishin W.,  Consolmagno G.~J.,  Mogren K.,
  Vilas F.,  2007, \mn@doi [The Astronomical Journal] {10.1086/510134}, 133,
  526

\bibitem[\protect\citeauthoryear{{Tegler}, {Grundy}, {Vilas}, {Romanishin},
  {Cornelison}  \& {Consolmagno}}{{Tegler} et~al.}{2008}]{tegler2008}
{Tegler} S.~C.,  {Grundy} W.~M.,  {Vilas} F.,  {Romanishin} W.,  {Cornelison}
  D.~M.,   {Consolmagno} G.~J.,  2008, \mn@doi [\icarus]
  {10.1016/j.icarus.2007.12.015}, \href
  {https://ui.adsabs.harvard.edu/abs/2008Icar..195..844T} {195, 844}

\bibitem[\protect\citeauthoryear{Young, Braga-Ribas  \& Johnson}{Young
  et~al.}{2020}]{Young2020}
Young L.~A.,  Braga-Ribas F.,   Johnson R.~E.,  2020, in Prialnik D.,  Barucci
  M.~A.,   Young L.~A.,  eds, , The Trans-Neptunian Solar System.
Elsevier, pp 127 -- 151,
  \mn@doi{https://doi.org/10.1016/B978-0-12-816490-7.00006-0}, \url
  {http://www.sciencedirect.com/science/article/pii/B9780128164907000060}

\makeatother
\end{thebibliography}








\bsp	
\label{lastpage}

\end{document}